\documentclass[twocolumn,showpacs,preprintnumbers,amsmath,amssymb,prl,superscriptaddress]{revtex4-1}
\usepackage{amsmath,amssymb,amsthm}
\usepackage{graphicx}
\usepackage{dcolumn}% Align table columns on decimal point

\newcommand {\beq}{\begin{eqnarray}}
\newcommand {\eeq}{\end{eqnarray}}

\begin{document}
%%%%%%%%%%%%%%%%%%%%%%%%%%%%%%%%%%%%%%%%%%%%%%%%%%%%%%%%%%%%%%%%%%%%%%%%
%%%%%%%%%%%%%%%%%%%%%%%%%%%%%%%%%%%%%%%%%%%%%%%%%%%%%%%%%%%%%%%%%%%%%%%%

\preprint{CALT-68-2810, IPMU10-0201, SCIPP 
10/20}

\title{Spatially Modulated Phase in Holographic Quark-Gluon Plasma}% Force line breaks with \\

\author{Hirosi Ooguri}

\affiliation{California Institute of Technology, 452-48, Pasadena, CA 91125, USA}
\affiliation{Institute for the Physics and Mathematics of the Universe,
University of Tokyo, Kashiwa, Chiba 277-8586, Japan}

%--
\author{Chang-Soon Park}

\affiliation{Santa Cruz Institute for Particle Physics 
and Department of Physics, University of California, Santa Cruz, CA 95064, USA}

%\date{\today}% It is always \today, today,
             %  but any date may be explicitly specified

%-----------------------------------------
\begin{abstract}
We present a string theory construction of a gravity
dual of a spatially modulated phase. In our earlier work,
we showed that the Chern-Simons term in the 5-dimensional
Maxwell theory destabilizes the Reissner-Nordstr\"om black 
holes in anti-de Sitter space if the Chern-Simons coupling 
is sufficiently high. In this paper, we show that a similar 
instability is realized on the worldvolume of 8-branes in 
the Sakai-Sugimoto model in the quark-gluon plasma phase. 
We also construct and analyze a non-linear solution describing 
the end-point of the transition. Our result suggests a new spatially 
modulated phase in quark-gluon plasma when the baryon density 
is above $0.8 N_f$ ${\rm fm}^{-3}$ at temperature 150 MeV.

\end{abstract}

% PACS, the Physics and Astronomy
                             % Classification Scheme.
%\keywords{Suggested keywords}%Use showkeys class option if keyword
                              %display desired
\maketitle

%%%%%%%%%%%%%%%%%%%%%%%%%%%%%%%%%%%%%%%%%%%%%%%%%%%%
\section{Introduction}

The 5-dimensional Maxwell theory with the Chern-Simons term is 
tachyonic in the presence of a constant electric field \cite{Nakamura}.
The tachyonic modes with non-zero spatial momenta 
can destabilize the Reissner-Nordstr\"om black holes in 5-dimensional 
anti-de Sitter space ($AdS_5$) if the Chern-Simons coupling is larger
than a certain critical value. If it happens, the holographically dual 
quantum field theory in (3+1) dimensions encounters a spatially modulated phase
transition. In \cite{Park}, we constructed and analyzed 
non-linear solutions in the bulk
which would describe the end-point of the 
 phase transition.

However, an explicit realization of such an instability in superstring
theory has been missing. For example, the Chern-Simons coupling of the 
minimal gauged supergravity in 5 dimensions is $\alpha = 1/2\sqrt{3} = 0.2887...$ 
which is slightly lower than the critical value $\alpha_{{\rm crit}} = 0.2896...$
for the instability of the extremal Reissner-Nordstr\"om black hole
\cite{Nakamura}. Similarly, the three-charge extremal black hole in the 
type IIB superstring theory on $AdS_5 \times S^5$ is found to be barely stable. 

In this paper, we show that such an instability is realized in the
quark-gluon plasma phase of the Sakai-Sugimoto model for QCD with
$N_f$ flavors of massless quarks \cite{Sakai}.
On the worldvolume of the 8-branes, there is a $U(N_f)$ gauge field, 
and its diagonal $U(1)$ part is dual to the quark number ($= N_c$ times the baryon number). 
The baryons are identified with instanton solutions on the worldvolume in this model \cite{Hata}. 
Worldvolume solutions representing QCD states with finite baryon density 
and temperature have been studied \cite{Sakai,Aharony,Bergman}.
See also \cite{densityone,densitytwo,densitythree} for related papers. 

Most of the solutions with finite baryon density are singular at the sources
of baryons charges, and it is not clear whether the supergravity approximation 
is applicable. One of the exceptional cases is the quark-gluon plasma phase,
where there is a smooth solution representing a finite baryon density 
configuration.

In the Sakai-Sugimoto model, the gluon degrees of freedom are realized on 
$N_c$ D4 branes compactified on a circle $S_c^1$ with supersymmetry breaking boundary
condition \cite{Witten}. At finite temperature, we compactify the Euclidean time
on another circle $S_T^1$, and the D4 brane world volume has the topology of $S_c^1 \times S_T^1 \times \mathbb{R}^3$.
In the confinement phase, $S_c^1$ is contractible in the bulk, and the topology of 
the bulk geometry is then $S^1_T \times \mathbb{R}^3 \times S^4$ times 
a disk bounded by $S_c^1$. Each 8-brane wraps the thermal $S^1_T \times S^4$ and 
is extended in $\mathbb{R}^3$. In this phase, the 8-brane starts as a D8 brane at 
a point on $S^1_c$, meanders in the bulk, and ends as a $\overline{{\rm D}8}$ brane 
at another point on $S^1_c$. 

In the deconfinement phase, the thermal $S^1_T$ becomes 
contractible in the bulk geometry \cite{Witten}. Depending on the relative
locations of the 8-branes, the chiral symmetry restoration
takes place at or above the deconfinement temperature \cite{Sakai,Aharony,Bergman}. Above the 
chiral symmetry restoration temperature, D8 and $\overline{{\rm D}8}$ branes become
separated, and each of them has the topology of a disk bounded by $S_T^1$ times $S^4$. 
This describes a holographic dual of the quark-gluon plasma in this model. 
In this phase, it is possible to construct a solution with finite baryon density that is smooth 
everywhere on the worldvolume, as we will discuss below. In this paper, we will
focus on this case. 

The dynamics of the 8-brane worldvolume is described by the Dirac-Born-Infeld (DBI) action
with the Chern-Simons term. We show that there is a critical baryon density above 
which the brane configuration becomes unstable by tachyonic modes carrying non-zero 
momenta. To understand the nature of the phase transition, we construct a solution 
to the full non-linear equations. Though the solution
carries non-zero momentum, its energy is lower than that of the original 
configuration which is spatially homogeneous. This suggests a spatially modulated phase with a baryon density wave. 

A holographic dual of a baryon density wave was discussed in a phenomenological 
model in \cite{Domokos}. The instability of the Sakai-Sugimoto model has been 
studied earlier, for example in \cite{Chuang}, but not in the chiral symmetric 
phase. To our knowledge, it has not been shown whether the Chern-Simons coupling on
the worldvolume theory on the 8-branes is large enough to trigger the spatially 
modulated phase transition. In this paper, we will show that the Chern-Simons coupling 
is 3 times the critical value required for the instability. 

\section{Instability of homogeneous solution}

The bulk geometry above the deconfining temperature is given by \cite{Witten},
\begin{equation}\label{actionone}
\begin{split}
ds^2 =&\left(\frac{U}{R}\right)^{\frac 3 2} \left(-f(U) d{X_0}^2 + d\vec {X}^2 + d{X_4}^2\right) \\
&+ \left( \frac{R}{U}\right)^{\frac 3 2} \left(\frac{dU^2}{f(U)} + U^2 d\Omega_4^2\right)\;,\\
e^{\phi} = &g_s \left(\frac U R\right)^{\frac 3 4}\;,~~
 F_4=dC_3 = \frac{3 N_c}{4\pi} \epsilon_4\;,
\end{split}
\end{equation}
where $f(U)=1-U_T^3/U^3$, $R^3 = \pi g_s N_c l_s^3$, $\epsilon_4$ is the volume form 
of a unit $S^4$, and $d\Omega_4^2$ is a metric for a unit four-sphere.
The temperature $T$ is $\frac 3 {4\pi} U_T^{1/ 2}R^{-3/2}$.
It sets the periodicity of the $X_0$ direction (along the thermal circle $S_T^1$) in the Euclidean solution, 
while the period of the compact $X_4$ direction (along the $S_c^1$) 
is arbitrary. In the chiral symmetry restoration phase, each 8-brane is located 
at a constant $X_4$ \cite{Sakai,Aharony,Bergman}. The induced metric on the branes is 
\begin{equation}
\begin{split}
ds^2 =&\left(\frac{U}{R}\right)^{\frac 3 2} \left(-f(U) dX_0^2 + d\vec X^2\right) \\
&
+ \left( \frac{R}{U}\right)^{\frac 3 2} \left(\frac{dU^2}{f(U)} + U^2 d\Omega_4^2\right)\;.
\end{split}
\end{equation}

The D8 and $\overline{{\rm D}8}$ branes are separated in the chiral symmetric phase. For now, let us focus on 
the dynamics on the D8 branes. The DBI action on the D8 brane is given by
\begin{equation}\label{E:D8action}
S=-T_{D8} \int d^9 \sigma e^{-\phi} \sqrt{-\det(g_{\alpha\beta}+2\pi\alpha' {F}_{\alpha\beta})}+S_{CS}\;,
\end{equation}
where $T_{D8}=(2\pi)^{-8} l_s^{-9}$.
The Chern-Simons action is given by
\begin{equation}
S_{CS}=\frac{1}{48\pi^3} \int_{D8} F_4 \wedge \omega_5(A)\;,
\end{equation}
where $F_4=dC_3$ is the RR 4-form field which satisfies $\frac 1 {2\pi} \int_{S^4} F_4 = N_c$ 
in our convention and $\omega_5(A) = A\wedge F\wedge F$ is the Chern-Simons 5-form.

For our purpose, it is sufficient to turn on the $U(1)$ part of the gauge field on the worldvolume.
To the quadratic order, the $U(1)$ part does not couple to 
the $SU(N_f)$ part of the gauge field or fluctuations of the 8-brane in the transverse direction. 
Couplings to the bulk degrees of freedom are suppressed by $1/N_c$. 
To simplify our equations, we rescale the gauge field and the metric as $A=\frac{R^2}{2\pi\alpha'} \tilde A$
and $g_{\alpha\beta} = R^2 \tilde g_{\alpha\beta}$. We also rescale the coordinates as
\begin{equation}
u=\frac{U}{R}\;,~~ t=\frac{X_0}{R}\;,~~ \vec x = \frac{\vec X}{R}\;,~~ \tau = \frac{X_4}{R}\;.
\end{equation}
Assuming that the gauge field is constant on the $S^4$, we obtain an effective 5-dimensional action,
\begin{equation}\label{E:actionwithoutR}
\begin{split}
S/c=& - \int_{M_4\times\mathbb{R}}  dt d^3 x du\, u^{\frac 1 4} \sqrt{-\det(\tilde g_{\alpha\beta}+ {\tilde F}_{\alpha\beta})}\\
& +\alpha  \int_{M_4\times\mathbb{R}} dt d^3 x du \, \epsilon^{\mu_1\mu_2\mu_3\mu_4\mu_5} 
\tilde A_{\mu_1}\tilde F_{\mu_2\mu_3} \tilde F_{\mu_4\mu_5}\;.
\end{split}
\end{equation}
with the 5-dimensional metric,
\begin{equation}\label{E:gRescaled}
\begin{split}
ds^2 &=u^{\frac 3 2} \left(-f(u) dt^2 + d\vec x^2 +d \tau^2\right) + \frac{1}{u^{\frac 3 2} f(u)} du^2,\\
&f(u)= 1-\frac{u_T^3}{u^3}, ~~ u_T = \left( \frac {4\pi}3\right)^2 R^2 T^2\;.
\end{split}
\end{equation}
The Chern-Simons coupling $\alpha$ is fixed to be
\begin{equation}
\alpha = \frac{3}{4},
\end{equation}
and the factor $c$ is
\begin{equation}\label{factorc}
c = \frac{8\pi^2}{3} T_{D8} N_f g_s^{-1} R^9. 
\end{equation} 
Note that, modulo the overall factor $c$, the action (\ref{E:actionwithoutR}) depends only on $u_T$.

If the kinetic term for the gauge field were of the Maxwell form $\tilde F^2$, the electric field strength could be 
made arbitrarily high by raising the baryon density, and any non-zero value of the Chern-Simons coupling would 
induce an instability of the type discovered in \cite{Nakamura}. With the DBI action, there is an upper bound for the field
strength, and it requires a more careful analysis to determine whether the instability takes place. 

Let us consider a 
background configuration with non-zero $\tilde A_0 = \tilde A_0(u)$. The equation of motion gives,
\begin{equation}\label{E:backgroundE}
\tilde E(u) = \frac{\tilde \rho}{\sqrt{\tilde \rho^2 +u^5}}\;,
\end{equation}
where $\tilde E=-\tilde F_{tu}=\partial_u \tilde A_0$. 
The integration constant $\tilde \rho$ will be identified as a rescaled value of
the quark density $\rho$ ($=N_c$ times the baryon density). As advertised in the introduction, this solution with finite
quark density is
regular everywhere on the brane.
Note that 
$\tilde \rho \rightarrow \infty$ gives $\tilde E \rightarrow 1$.
 We choose the gauge potential $\tilde A_0(u)$ so that it vanishes on the horizon.
Since the chemical potential $\tilde \mu$ is given by the asymptotic value of $\tilde A_0$ at $u\rightarrow \infty$, we have
\begin{equation}
\tilde \mu=\tilde A_0(u=\infty) = \int^{\infty}_{u_T} du \frac{\tilde \rho}{\sqrt{\tilde \rho^2+u^5}}\;.
\end{equation}

Let us
 perturb it as $\tilde F \rightarrow \tilde F + \delta \tilde F$. To find an onset of the instability, we look for a
static normalizable solution in the linearized equation for $\delta \tilde F$,
\begin{equation}
\begin{split}
&\partial_u\left[ \frac{u^{\frac  5 2}f(u)}{\sqrt{1-\tilde E(u)^2}} \delta \tilde F_{u i}\right] 
-  u^{-\frac 1 2} \sqrt{1-\tilde E(u)^2} \partial_j \delta \tilde F_{ij} \\
&~~~~~~ - 2 \alpha \epsilon_{ijk} 
\tilde E(u) \delta \tilde F_{jk} =0\;
\end{split}
\end{equation}
By applying $\epsilon_{ijk} \partial_j$ to the $k$-th component of the above equation and 
using the expression \eqref{E:backgroundE} for $E(u)$, 
we find
\begin{equation}\label{E:fluctuationEquation}
\begin{split}
&\partial_u \left[f(u) \sqrt{\tilde \rho^2+ u^5}\partial_u  \delta \tilde F_i\right]
+\frac{u^2}{\sqrt{\tilde \rho^2+ u^5}} \partial_j\partial_j \delta \tilde F_i\\
&~~~~~~ -4\alpha \frac{\tilde \rho}{\sqrt{\tilde  \rho^2+ u^5}} \epsilon_{ijk} \partial_j \delta 
\tilde F_k =0\;,
\end{split}
\end{equation}
where $\delta\tilde F_i = \frac 1 2 \epsilon_{ijk} \delta \tilde F_{jk}$.
In the Fourier mode $\delta F_i = v_i e^{-i k_j x^j} \phi(u)$ with
the polarization $v_i$ being an eigenvector of $\epsilon_{ijk} k_j$ with an eigenvalue 
$i k = i | \vec{ {k}} |$ (we can also consider $-i k$ for an eigenvalue with 
the same result), $\phi(u)$ obeys a second order ordinary differential equation,
\begin{equation}
\label{ODE}
\left[ - \frac{d}{d u}
 f(u) \sqrt{\tilde \rho^2 + u^5} \frac{d}{du} + \frac{-4 \alpha \tilde \rho k + u^2 k^2 }{\sqrt{
 \tilde \rho^2 + u^5}} \right] \phi(u) = 0. 
 \end{equation}
At the horizon $u= u_T$, we use the in-going boundary condition for static waves.

\begin{figure}[ht]
\centering
\includegraphics[width=7cm]{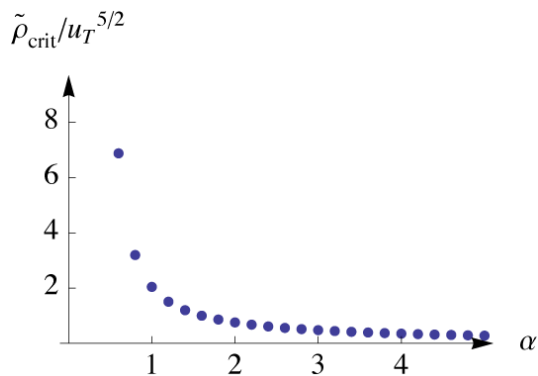}
\caption{The critical quark density $\tilde \rho$ as a function of the Chern-Simons coupling $\alpha$.}
\label{fig:alphadlinear}
\end{figure}

We solved the linearized equation (\ref{ODE}) numerically for general values of the Chern-Simons coupling $\alpha$.
For each value of the Chern-Simons coupling $\alpha> 1/4$, we found a critical value of $\tilde \rho$ 
above which the instability takes place.
Figure 1 depicts the critical quark density $\tilde \rho_{{\rm crit}}$ as a function of $\alpha$. We note that
$\tilde \rho_{{\rm crit}}$ diverges as $\alpha \rightarrow 1/4$. 

We can also show analytically that $\alpha = 1/4$ is the limiting value of the Chern-Simons coupling.
Let us rescale variables as 
\begin{equation}
\bar u = \tilde \rho^{-\frac{2}{5}} u, ~~ \bar k = \tilde \rho^{-\frac{1}{5}} k, 
\end{equation}
and take the limit $\tilde \rho \rightarrow \infty$ in the linearized equation (\ref{ODE}). We find
\begin{equation}
\label{SL}
\left[ - \frac{d}{d\bar u} \sqrt{1 +\bar u^5} \frac{d}{d\bar u}
 + \frac{-4 \alpha \bar k + \bar u^2 \bar k^2}{\sqrt{1 + \bar u^5}} \right] \tilde \phi(\bar u) = 0. 
\end{equation}
We have verified that a solution to this equation approaches the solution to (\ref{ODE}) 
in the sense of the ${\cal L}^2$ measure.
 From the numerical evaluation of (\ref{SL}), we find that the momentum $\bar k$ with 
 non-trivial normalizable solutions tends to infinity as we take $\tilde \rho \rightarrow \infty$ and
 $\alpha$ approaches the limiting value. Anticipating this, we take $\bar k  \rightarrow \infty$ in
 (\ref{SL}) while keeping $v=\sqrt{\bar k} \bar u$ and obtain,
 \begin{equation}
 \left( - \frac{d^2}{d v^2} 
 - 4\alpha + v^2 \right) \tilde \phi(\bar u) = 0. 
  \end{equation}
  This can be solved by the harmonic oscillator ground state $\phi(v) = e^{-v^2/2}$ with 
  $\alpha = 1/4$.

In the quark-gluon plasma phase, the Chern-Simons coupling on the worldvolume theory is 
$\alpha = 3/4$ and is above the limiting value of $1/4$. At this value of $\alpha$, 
the critical quark density is
numerically evaluated as
\begin{equation}\label{criticaltilded}
\tilde \rho_{{\rm crit}}= 3.714\ u_T^{\frac{5}{2}}.
\end{equation}

Let us express the critical quark density in the original set of variables. The quark density $\rho$ is defined by a variation
of the Lagrangian density with respect to $E = \partial_u A_0$. In the above, we rescale the action 
by the factor $c$ given by (\ref{factorc}) and the gauge field is rescaled as $A = 
\frac{R^2}{2\pi \alpha'} \tilde A$. We should also remember that we rescaled our spacetime coordinates by 
$R$. The physical quark density $\rho$ is then related to $\tilde \rho$ discussed in the above as
\begin{equation}
 \rho = c \left(\frac{R^2}{2\pi\alpha'}\right)^{-1} \frac{\tilde \rho}{R^3} 
 = \frac{2}{3(2\pi)^5} \frac{N_f}{g_s} \frac{R^4}{l_s^7} \tilde \rho .
 \end{equation}
Substituting (\ref{criticaltilded}) into this, the critical quark density at $\alpha = 3/4$ is given as
\begin{equation}\label{E:rhocrit}
\rho_{{\rm crit}} = c_0 N_f N_c (g_s N_c l_s)^2 T^5 ,  
\end{equation}
where $c_0 = 3.714 (2/3)^6 \pi^3 \approx 10$. 

It is important to make sure that we can ignore backreaction of this quark density to the bulk geometry. 
One way to see this is to note that the critical baryon density is given 
by dividing the quark density $\rho_{{\rm critial}}$ by $N_c$. The result is proportional to $N_f (g_s N_c)^2 T^5$, but it
does not have any power of $N_c$. Since the baryons can be thought of as D4 branes wrapping $S^4$ \cite{Wittentwo,GO},
their backreaction becomes significant only when their density scales as $N_c$ or more.
Thus, the backreaction
is negligible provided $N_f \ll N_c$. 
Another way to see this is to evaluate the energy density due to the electric field using
the action (\ref{E:actionwithoutR}) and show that it is proportional to $N_f/g_s$ times some power of
%the 't Hooft coupling
$g_s N_c$. This is the same scaling behavior as the tension of the $N_f$ 8-branes.

It is an interesting exercise to express the critical density in terms of QCD quantities. 
The string parameters $g_s$ and $l_s$ are related to the Yang-Mills
coupling $g_{YM}$ and the Kaluza-Klein scale $M_{KK}$ for the compactification circle $S^1_c$
as $g_{YM}^2=4\pi^2 g_s l_s / L$ and $M_{KK}=2\pi/L$, where $L$ is the circumference of $S_c^1$ \cite{towards}.
The critical baryon density can then be written as, 
\begin{equation}\label{criticaldensity}
\frac{\rho_{{\rm crit}}}{N_c} = \frac{c_0 N_f}{4\pi^2} \frac{\lambda^2}{M_{KK}^2} T^5,
\end{equation}
where $\lambda=g_{YM}^2 N_c$ is the 't Hooft coupling.
The constants $M_{KK}$ and $\lambda$ can be determined by fitting, for example, with 
the pion decay constant and the mass of the $\rho$-meson, as $M_{KK} =$ 949 MeV and
$\lambda = g_{YM}^2 N_c =$ 16.6 \cite{Sakai}. The deconfinement temperature, where
the thermal cycle $S_T^1$ becomes contractible, is at $M_{KK}/2\pi =$ 151 MeV.
Interestingly, this turns out to be close the critical temperature expected for 
the quark-gluon plasma. If we substitute $T = 150$ MeV in (\ref{criticaldensity}), 
for example, the critical baryon density comes out as,
\begin{equation}
\frac{\rho_{{\rm crit}}}{N_c} \approx 0.8 N_f \ {\rm fm}^{-3}.
\end{equation}
For $N_f=2$, this is about 10 times the nucleon density in atomic nuclei. 

\section{End-point of the phase transition}

Let us study the full non-linear equations to identify the end-point of the instability.
Following \cite{Park}, we make the following ansatz:
\begin{equation}\label{E:Aansatz}
\begin{split}
\tilde A_t &= a(u)\\
\tilde A_x+ i \tilde A_y &= h(u) e^{- i k z}\;,
\end{split}
\end{equation}
with the other gauge field components vanishing. 
They obey 
\begin{eqnarray}
\label{E:EOMforNonLinear}
&\partial_u \left[\frac{u\sqrt{u^3+k^2 h(u)^2} a'(u)}{\sqrt{1-a'(u)^2 + f(u) h'(u)^2}}\right] + 4\alpha k h(u) h'(u)=0 \nonumber \\
&\partial_u \left[\frac{u f(u) \sqrt{u^3+k^2 h(u)^2} h'(u)}{\sqrt{1-a'(u)^2 + f(u) h'(u)^2}}\right] 
+4\alpha k a'(u) h(u)\nonumber \\
&~~~~ -\frac{k^2 u h(u) \sqrt{1-a'(u)^2 + f(u) h'(u)^2}}{\sqrt{u^3+ k^2 h(u)^2}}=0.
\end{eqnarray}
We assume that the embedding coordinate $\tau$ is constant, and this assumption is consistent with the equations of motion. 
The first equation can be integrated easily, and gives us the quark density analogously to the homogeneous solution.
\begin{equation}\label{E:EOMtilderho}
\frac{u\sqrt{u^3+k^2 h(u)^2} a'(u)}{\sqrt{1-a'(u)^2 + f(u) h'(u)^2}}+2\alpha k h(u)^2 = \tilde \rho\;.
\end{equation}
Using this expression, the second equation becomes
\begin{equation}\label{E:nonLinearFinalEquation}
\begin{split}
 K(u) \partial_u \left( K(u) f(u)h'(u) \right) - k^2 u^2 h(u)& \\
+ 4 k \alpha h(u) (\tilde \rho-2 k \alpha h(u)^2) &=0\;,
\end{split}
\end{equation}
where
\begin{equation}
K(u) = \sqrt{\frac{\tilde \rho^2+u^5 + k h(u)^2 (k u^2 - 4 \tilde \rho \alpha + 4 k \alpha^2 h(u)^2}{1+f(u) h'(u)^2}}\;.
\end{equation}

\begin{figure}[ht]
\centering
\includegraphics[width=7cm]{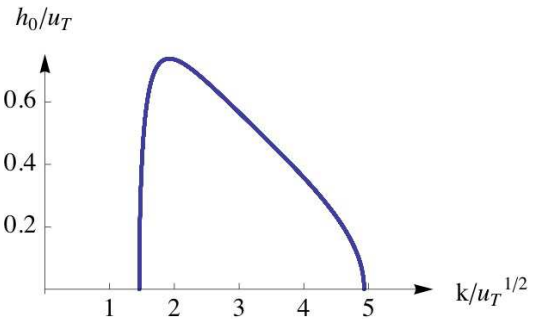}
\caption{Static normalizable 
solutions exist along this curve in the $h_0$-$k$ plane, when $\alpha=\frac 3 4$ and $\tilde \rho=5 u_T^{5/2}$. 
The minimum free energy occurs at $k= 2.35 u_T^{1/2}$, which is slightly larger than $k$ for the maximum
value of $h_0$.}\label{fig:h0ksample}
\end{figure}

The equation (\ref{E:nonLinearFinalEquation}) can be solved numerically. 
Figure \ref{fig:h0ksample} shows the range of momenta where non-linear static solutions exist
for a sample case of $\tilde \rho=5 u_T^{5/2}$. The vertical axis $h_0$ is an initial value of $h(u)$: $h_0=h(u_T)$.
Since we have  a family of solutions parametrized by
the momentum $k$, we look for the one which minimizes the free energy density
$\mathcal{F}$, given by
\begin{equation}
\mathcal{F}(\rho) = \mu \rho + \int du\, \mathcal{L}_E ,
 \end{equation}
where $\mathcal{L}_E$ is the DBI Lagrangian plus the Chern-Simons term.
Note that the free energy $\mathcal{F}$ is a function of $\rho$, and not $\mu$.

We have identified the momentum with the lowest value of the free energy, and 
the expectation value of the current operator  $\langle \tilde J\rangle$ dual to $h(u)$ can be read off 
from the asymptotic behavior of the normalizable solutions.
That is, the current corresponding to \eqref{E:Aansatz} has $x$ and $y$ components $\tilde J_x + i \tilde J_y = \tilde J e^{-i k z}$.

At the critical density given by \eqref{E:rhocrit}, the instability begins to occur at the momentum 
$k=2.39 u_T^{1/2}$, which in the original coordinates is given by $k / R \approx 10T$.
%Thus, the momentum at the peak of Fig. 2 is $2 u_T^{1/2} R^{-1} = \frac{8\pi}{3} T$. 
If we set $T=150$ MeV, the momentum is about $1.5$ GeV, and the corresponding wave length is $0.8$ fm.   

\begin{figure}[ht]
\centering
\includegraphics[width=7cm]{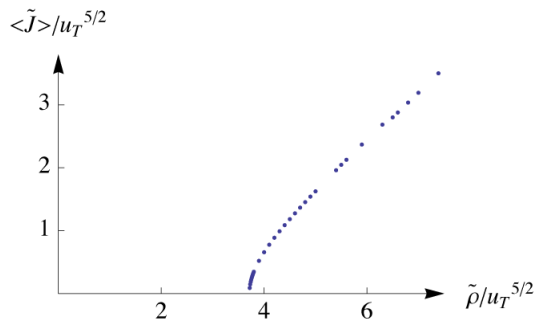}
\caption{The expectation value of the dual current operator $\tilde{\left<J\right>}$ as a function of $\tilde \rho$
at  $\alpha=\frac 3 4$.}\label{fig:nonlineardvsJ}
\end{figure}

Figure \ref{fig:nonlineardvsJ} shows the relation between $\tilde \rho$ and $\langle \tilde{J}\rangle$.
Note that there is a critical value of $\tilde \rho$ below which there is no spatially modulated solution.
In terms of the original coordinates in \eqref{actionone}, the dual current $\langle J \rangle$ is given by
\begin{equation}
\frac{\langle J\rangle}{N_c} = \frac{\pi}{4}\left(\frac 2 3\right)^6 \frac{\lambda^2}{M_{KK}^2} N_f T^5 
\cdot \frac{\langle\tilde J\rangle}{u_T^{5/2}}\;.
\end{equation}

So far, we have focused on the dynamics on the D8 brane worldvolume.
The analysis on the $\overline{{\rm D}8}$ branes is identical except
that the Chern-Simons coupling has the opposite sign due to the CPT invariance. 
There are separate gauge fields $A_L$ and $A_R$ on the D8 and $\overline{{\rm D}8}$ branes,
respectively, and they cause the instability above the critical charge density.
The baryon vector current is dual to $(A_L+A_R)$ and the axial current is dual to $(A_L - A_R)$.
The baryon charge density turns on the same amount of chemical potentials for both
$A_L$ and $A_R$. Above the critical baryon density, the instability will take place on both
branes, each of which will settle in a configuration carrying a momentum of the size 
$|k|/R \approx 10 T$. 
However, directions of the momenta on the D8 and $\overline{{\rm D}8}$ branes can be different,
and the currents $J_L$ and $J_R$ dual to $A_L$ and $A_R$ can carry momenta in different
directions. Thus, in the spatially modulated phase, both vector and axial baryon currents 
are generated on the boundary.

%%%%%%%%%%%%%%%%%%%%%%%%%%%%%%%%%%%%%%%%%
\section*{Acknowledgments}

We would like to thank Makoto Natsuume, Dam Son, Shigeki Sugimoto, and Chen-Pin Yeh 
for stimulating discussions. H.O. thanks the Aspen
Center for Physics and the Simons Center for Geometry and Physics, 
and C.S.P. thanks the IPMU for their hospitalities,
where parts of this work were carried out. 

This work is supported in part 
by DOE grant DE-FG03-92-ER40701 and by the 
World Premier International Research Center Initiative of MEXT.
H.O. is also supported in part by a Grant-in-Aid for Scientific 
Research (C) 20540256 of JSPS. 
C.S.P. is also supported in part by DOE grant DE-FG02-04ER41286.

%%%%%%%%%%%%%%%%%%%%%%%%%%%%%%%%%%%%%%%%%%%%%%%%%%%%%%%%%%%%%%%%%%%%%%%%%%%%%%%%%%

\end{document}